\documentclass{aastex} 
\usepackage{emulateapj5}

\newcommand{\myemail}{hwlee@galaxy.yonsei.ac.kr} 
  
\slugcomment{To appear in ApJ Letters}
 
\shorttitle{H$\alpha$ Wings in Symbiotic Stars}
\shortauthors{Lee}
  
\begin{document} 
   
\title{Raman Scattering Wings of H$\alpha$ in Symbiotic Stars}

\author{Hee-Won Lee}
\affil{Department of Astronomy, Yonsei University \\
       Seoul, Korea}
\email{\myemail}

\begin{abstract}
Nussbaumer et al. (1989) proposed that broad H$\alpha$ wings 
can be formed through Raman scattering of Ly$\beta$ photons, 
and in this Letter we argue that the H$\alpha$ wings prevalently seen
in symbiotic stars may be indeed formed in this way. 
Assuming a flat incident UV radiation around Ly$\beta$, we generate
template wing profiles around H$\alpha$ that are formed through 
Raman scattering in a plane-parallel H~I region.  We
perform profile fitting analyses to show that the template wing profiles 
are in excellent agreement with the observed ones that are provided
by van Winckel et al. (1993) and Ivison et al. (1994). The wing flux is
determined by the scattering H~I column density and the incident Ly$\beta$
flux strength and profile. From our profile analysis it is proposed that the
Raman scattering component may be identified with the  
neutral envelope with a column density ranging $10^{18-20} {\rm\ cm^{-2}}$
that surrounds the binary system. 
We briefly discuss alternative candidates for the wing formation mechanism
and observational implications of Raman
scattering in symbiotic stars and in other astronomical objects including 
planetary nebulae, post AGB stars and active galactic nuclei. 
\end{abstract} 
\keywords{line: formation --- line: profiles --- radiative transfer --- 
scattering --- (stars) binaries  : symbiotic} 

\section{Introduction}

A long standing puzzle regarding the identification of the so called
symbiotic bands around 6830 \AA\ and 7088 \AA\ occurring 
in about half of symbiotic stars has been solved by Schmid (1989),
who proposed that they are the O~VI 1032, 1038 doublet features that
are Raman scattered by atomic hydrogen. These Raman-scattered
lines possess very broad profiles because of the broadening
enhancement by a factor of about 6.4 due to the scattering incoherency, and
they exhibit rich structures including
multiple peaks and high degrees of polarization often accompanied
by a position angle flip in the red wing parts (Harries \& Howarth
1996, Schmid \& Schild 1994, Schmid et al. 2000). 

Colliding winds models are often invoked to explain the abundant
structures shown in the Raman scattered lines (e.g. Girard \& Willson 1987).
Another point of view was presented by Lee \& Park (1999), who argued that
an accretion disk type emission model may be more adequate for explanations
of the spectropolarimetric observational data. The Raman scattered lines
should be important in that they provide unique information that
can be obtained from a mirror located on the binary axis connecting 
the giant and the dwarf.

Nussbaumer et al. (1989) emphasized the importance of Raman scattering
in astrophysics, and proposed that it may play an important role in the 
formation of broad H$\alpha$ wings from incident Ly$\beta$ photons. 
Lee \& Hyung (2000, hereafter LH) computed the template wing profiles
that are formed through Raman scattering
in a plane-parallel H~I region with various H~I column densities, and
applied their analysis to the young and compact planetary nebula IC~4997.
The excellent agreement between the template profile with the observed
one supports the hypothesis that the broad H$\alpha$ wings in
IC~4997 are formed through Raman scattering in a neutral
envelope with a H~I column density $N_{HI} \sim 10^{20}{\rm\ cm^{-2}}$,
which is also consistent with the H~I 21 cm absorption observation 
of IC~4997 (Altschuler et al. 1986).

The neutral envelope found in IC~4997 is an important element to
study the mass loss process of the central star and we turn our
attention to find similar evidence in symbiotic stars.
Van Winckel et al. (1993) have compiled high resolution H$\alpha$ line
profiles of symbiotic stars in the southern hemisphere and the northern
counterparts have been done by Ivison et al. (1994). It was shown that
a typical H$\alpha$ profile exhibits a dip in the center part and
broad wings that often cover the narrow [NII] 6548, 6584 lines.
Despite many studies on the H$\alpha$ profiles of symbiotic stars,
it is not still clear whether they consist of multiple emission components
or the center part is absorbed. A recent study by Schwank et al. (1997) 
shows that the central dip may represent the absorption
and that the wing parts have a different origin.
In this Letter, we perform profile fitting analyses to propose that 
Raman scattering may be responsible for the H$\alpha$ broad wings 
observed in symbiotic stars.

\section{Profile Fitting Analysis}
\subsection{Assumptions}

Most H$\alpha$ wings
in symbiotic stars show velocity widths $\sim 1000 {\rm\
km\ s^{-1}}$.  If they are formed through Raman scattering, then
this implies that the kinematics of the emission region is characterized
by a typical velocity of order $150{\rm\ km\ s^{-1}}$, considering the 
Doppler enhancement factor $\lambda_{H\alpha} /\lambda_{Ly\beta}=6.4$. 
The detailed kinematic structure of the
emission region is not certain at the moment, and this velocity
scale exceeds by an order of magnitude the wind velocity associated with 
the mass loss of the giant companion in a symbiotic system. However, the Raman
scattered features at 6830 \AA\ and 7088 \AA\ have typical widths of
$\ge 20{\rm\ \AA}$, which also requires a similar velocity scale of about
$100{\rm\ km\ s^{-1}}$ in the O~VI emission region. From the ORFEUS 
observations, O~VI $\lambda\lambda$ 1032, 1038 lines 
with FWHM up to $100{\rm\ km\ s^{-1}}$ have been observed
in several symbiotic systems (Schmid et al. 1999). One plausible
suggestion is that the UV lines are formed in an accretion disk
type emission region that extends a sub-AU scale around the 
white dwarf component (Lee \& Park 1999, Morris 1987).

In the profile analysis performed on the H$\alpha$ wings in IC~4997 by LH,  
it was assumed that there exists a dense emission region characterized 
by an electron density $n_e \sim 10^{9-10}{\rm\ cm^{-3}}$ and
the photoionization code `CLOUDY' was used to show that a
sufficient number of Ly$\beta$ photons are available for Raman
scattering in the neutral envelope that is supposed to cover
the central emission line region (Ferland 1996). Noting the similarity of the
overall H$\alpha$ profile of IC~4997 and those of symbiotic stars
compiled by van Winckel et al. (1993) and Ivison et al. (1994), we
assume that Ly$\beta$ flux is available with a smooth
profile and concentrate on the wing formation through Raman
scattering.  

In order to generate H$\alpha$ wing profiles that may extend
in the range 6540 \AA\ through 6580 \AA, we assume that the incident 
Ly$\beta$ flux has a top-hat profile with a full width $300{\rm\ km\ s^{-1}}$.
This velocity scale is somewhat larger than that associated with the O~VI 
emission lines observed in ORFEUS. It is pointed out here that if we use
a narrow profile as the incident Ly$\beta$ flux such as a Gaussian with 
FWHM$\la 100{\rm\ km\ s^{-1}}$, then the Raman scattering will not
produce the broad H$\alpha$ wings  observed in symbiotic stars.

The neutral
envelope may possess an expansion velocity of order $10{\rm \ km\ s^{-1}}$
or higher, but this value is much smaller than the total wing width. The bulk
velocity of the scattering region also shifts the center of the
wing profiles with respect to the center of the core emission part.
However, in this Letter, we neglect these small velocity shifts and
concentrate on the wing formation process in a medium that is assumed to 
be static.

\subsection{Template H$\alpha$ Wing Profiles}

In order to prepare the template profiles, LH
computed the Raman conversion factor $C_{R}(\lambda)$ which is defined
as the ratio of the number of Raman-scattered and emergent H$\alpha$ photons 
to that of the incident Ly$\beta$ photons.  
In the wavelength interval $6540{\rm\ \AA\ }\le \lambda_o \le 6580{\rm\ \AA}$
the branching ratio of the Raman scattering with respect to the
Rayleigh scattering is approximately given by the relation
\begin{equation}
\sigma_{Ram}/\sigma_{Ray} \simeq 
0.1342 +11.06\ \Delta\lambda_i/\lambda_{Ly\beta}, 
\end{equation}
where $\Delta\lambda_i \equiv \lambda_i - \lambda_{Ly\beta}$ is the
difference of the wavelength $\lambda_i$ of the incident photon and 
that of the Ly$\beta$ line center (see Lee \& Yun 1998).

The wing profile is obtained in the optically thin limit, where almost
all the incident Ly$\beta$ photons are scattered no more than once. 
In this limit, from the Kramers-Heisenberg formula, the total scattering
cross section that is the sum of the Rayleigh and Raman scattering cross
sections is approximated by the far Lorentzian wing part, and therefore
\begin{equation}
\sigma_{tot} \propto \Delta\lambda_i^{-2}+O(\Delta\lambda_i^0) .
\end{equation}
Here, $O(\Delta\lambda_i^0)$ is the remainder term dominated by the
zeroth order of the wavelength shift (e.g. Sakurai 1967),
and we neglected the damping constant associated with the life time
of the $2p$ state of hydrogen atom.

Therefore, the wing profile is given by the relation
\begin{equation}
f_{\lambda} \propto 0.1342/ \Delta\lambda_i^{2}
+11.06 /(\Delta\lambda_i\lambda_{Ly\beta}) + \cdots.
\end{equation}
This implies that the wing profile is approximated to the first
order by the curve proportional to $f(\Delta v) = \Delta v^{-2}$
with the flat incident Ly$\beta$ profile. 

In the opposite limit where the scattering optical depth is much
larger than 1, $C_R$ converges to
a constant value that depends sensitively on the branching ratio 
$\sigma_{Ram}/\sigma_{Ray}$. 
Schmid (1996) showed that a Monte Carlo technique is useful to compute 
the efficiency of Raman scattering, and we use a Monte Carlo code to
find that $C_R$ reaches 0.6 in the optically
thick limit, which is consistent with the empirical formula proposed
by Lee \& Lee (1997).

\subsection{Results}

In Fig.~1, we compared the template profiles with the H$\alpha$ profiles 
of the symbiotic star SY Mus obtained by van Winckel et al. (1993). We
enlarged the profiles by 5 times to show the detailed wing profiles.  By
the thick dashed lines we represent the profiles proportional to
$f(\Delta v) = \Delta v^{-2}$, which correspond to the
envelope profiles in the optically thin limit and a flat
incident radiation. The dotted lines and
the long dashed lines show
the wing profiles that are generated using the Monte Carlo technique
adopted in the work of LH. The long dashed lines
represent the wing profiles obtained from the scattering region
of H~I column density $N_{HI} = 5\times 10^{19} {\rm cm^{-2}}$ and the
dotted lines from $N_{HI}= 10^{19}{\rm cm^{-2}}$. Notice that the
wing strength is determined by the product of the Ly$\beta$ flux
and the scattering column density, so that the solid blue lines are
obtained from 5 times stronger incident Ly$\beta$ flux than in the
case of the dotted blue lines.

In Fig.~2, we show the result of our fitting analysis to 16
symbiotic stars selected from the atlas of H$\alpha$ spectra
compiled by van Winckel et al. (1993) and Ivison et al. (1994).
In the present work we selected objects with conspicuous wings, and
no particular selection criterion was applied. We note that
most H$\alpha$ profiles not shown in Fig.~2 also showed similar
quality of agreement with our template profiles.
In Fig.~2, we used the Monte Carlo profile obtained from the scattering 
column of $N_{HI}= 10^{19}{\rm cm^{-2}}$ for simplicity of presentation.
There is an excellent agreement between the observed H$\alpha$ wing profiles
and the template wing profiles, which implies
that Raman scattering is definitely a possible mechanism for the formation
of the observed H$\alpha$ line wings in symbiotic stars.

\section{Discussion}

\subsection{Alternative Candidates for the H$\alpha$ Wing Formation Mechanism} 

We may consider other theoretical possibilities for the H$\alpha$ wing
formation mechanism. These include electron scattering, fast stellar winds
or outflows and H$\alpha$ damping wing scattering.
The electron scattering cross section
is wavelength independent and therefore similar wings may be found in
other emission lines including higher Balmer line series that may be
formed in the same emission region as H$\alpha$. Spectroscopy with higher 
signal to noise ratio will be necessary to distinguish electron scattering
wings from Raman wings.

If the H$\alpha$ wings in symbiotic stars represent the kinematics of
the ionized hydrogen, then it may imply the existence of fast stellar winds
or outflows. These outflows are often found in various objects including 
symbiotic stars, Wolf-Rayet stars and luminous blue variables. In particular, 
Schild, M\"urset \& Schmutz  (1996)
fitted the line wings of RW Hydrae with synthetic wind profiles and 
concluded that the wings are consistent with a wind feature. Wind emission
may be distinguished from the scattered features by polarimetry, which
is briefly discussed later.

According to Schwank et al. (1997) the core part of H$\alpha$
is formed mainly in the transition zone where ionized hydrogen
recombines. They show that the Balmer lines are optically thick and therefore
it is a possibility that the broad wings may be due to the H$\alpha$ damping
wings. Because the H$\alpha$ damping wings are also
characterized by the same dependence $\propto \Delta \lambda^{-2}$,
this process is worth further discussion.

However, in this case the wing photons are also generated in the
same region as the core photons and therefore we may expect that the
wing strength should be strongly correlated with that of the core part.
Schmutz et al.(1994) presented the H$\alpha$ profiles at various orbital
phases of the symbiotic star SY Mus. They pointed out that the
H$\alpha$ core part of this object shows regular variations 
both in the profiles and in the line intensity, whereas the wing part
remains constant. This implies that the wing part may be formed 
in a much more extended region than in the region responsible for the core
part.

It is notable that the Raman scattered
H$\alpha$ wings can be strongly polarized depending on
the scattering geometry. Because symbiotic stars usually exhibit
bipolar nebular morphology, which may represent an anisotropic
matter distribution around the central stellar region, we may expect that
neutral hydrogen is distributed in a similar way. 
In this case, the H$\alpha$ wings can be strongly polarized
with the polarization direction parallel or perpendicular to the polar axis. 
Furthermore, if the neutral envelope is expanding, then 
much stronger polarized flux will be obtained in the red wing part than in 
the blue wing part. 

A hint of this behavior is shown in the spectropolarimetric observation of
BI Crucis presented by Harries (1996), who proposed the accretion disk
reflection as the main mechanism for polarization generation. However, 
the overall spectropolarimetric behavior shown in BI Cru can be also
interpreted in the context of the Raman
scattering in an expanding neutral envelope, of which more detailed theoretical
study is under way (Bak \& Lee, in preparation). Further spectropolarimetric
observations of H$\alpha$ in symbiotic stars are expected to provide important
clues to the origin of the broad H$\alpha$ wings. 

At present, other mechanisms cannot be excluded as the origin 
of the H$\alpha$ broad wings in symbiotic stars. However, in view of
the current profile fitting analyses and
the existence of the Raman scattered O~VI lines in symbiotic stars,
we propose that the Raman scattering of Ly$\beta$ by atomic hydrogen 
may provide the most plausible explanation for
the H$\alpha$ wing formation.

\subsection{Observational Ramifications}

In the present work, the essential parameters are the strength of
the incident Ly$\beta$ and the scattering column density $N_{HI}$,
of which the product determines the wing flux profile and therefore
becomes the only free parameter in the profile fitting analysis.
In the fitting analysis performed on the planetary nebula
IC~4997 by LH, there exist H~I observations to obtain an 
independent measurement of $N_{HI}$ (see Altschuler et al. 1986). 
In the present work no such observations
seem to be available at this time, and similar H~I observations
may provide important information about the wing formation mechanisms.

From the strength of the Raman scattered features around 6830 \AA\ and 
7088 \AA, it is often proposed that they are scattered in a region
with $N_{HI} \ga 10^{22}{\rm cm^{-2}}$. It appears that these features
are formed deep inside the symbiotic system where the scattering region
is located near the giant component. However, since Ly$\beta$ line photons
have much larger scattering cross section than O~VI $\lambda\lambda$
1032, 1034 photons do, the H$\alpha$ wings may 
also be contributed significantly in a more extended region with a much 
lower $N_{HI}$ than the O~VI scattering region.  
Furthermore,
it is not apparent that the wings are formed in a scattering region
of a uniform density, because a composite profile from a number of
scattering regions with different $N_{HI}$ will produce a
similar profile with the dependence of Eq.~(3). More detailed information
may be obtained from high resolution far UV spectroscopy around Ly$\beta$,
which may be achieved by the FUSE mission (see also Schmid et al. 1999).

The recent report by Van de Steene, Wood, \& van Hoof (2000) shows that
broad H$\alpha$ wings are also present in a number of post AGB stars.
Balick (1989) also reported the existence of very broad wings around H$\alpha$
in the planetary nebula M2-9. It appears that the broad H$\alpha$ wings
are a common feature of post AGB stars, symbiotic stars and young
compact planetary nebulae, most of which are believed to be (or have been)
in the process of heavy mass loss giving rise to the formation
of a neutral envelope around the hot emission region.
As Nussbaumer et al. (1989) noted, the Raman scattering may
also operate in active galactic nuclei (AGN) to form broad H$\alpha$ wings. 
Romano et al. (1996) investigated the H$\alpha$ wing profiles of AGN
and found that they are well fitted by the relation $\Delta v^{-2}$.
In AGN the H$\alpha$ wing width often exceeds $10^4{\rm\ km\ s^{-1}}$
and we may expect to find a slight deviation from $\Delta\lambda^{-2}$
due to the contribution from the second term in Eq. (3), of which
more theoretical and observational work is needed to test this hypothesis.

\acknowledgements
The author is grateful to Yong-Ik Byun for helpful discussion. He also
thanks the referee Werner Schmutz, who suggested various important
points and helped improve the presentation of this paper.  HWL 
gratefully acknowledges support from the BK21 project initiated by
the Korea Ministry of Education.

\begin{figure}
\plotone{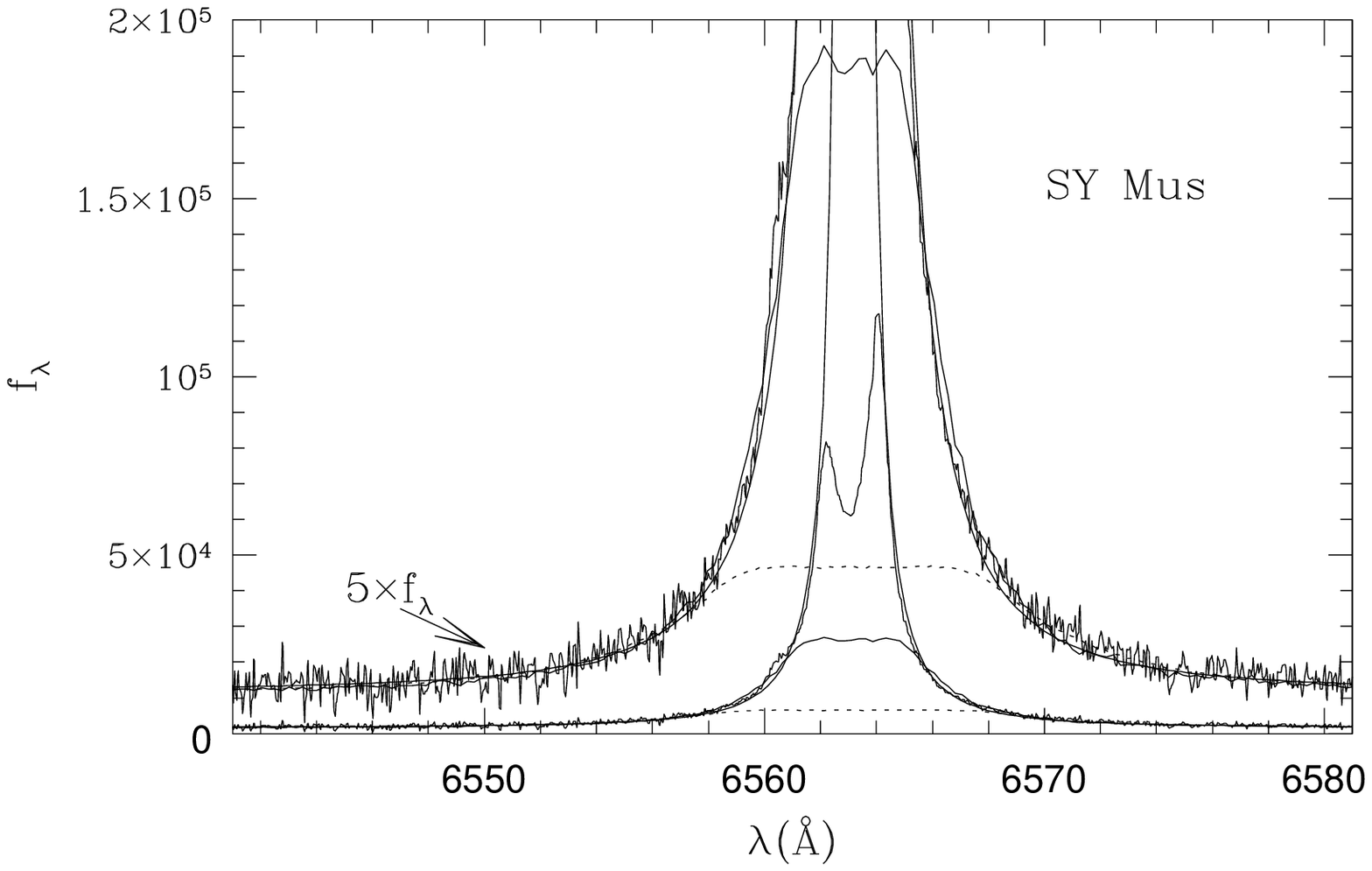}
\caption{
H$\alpha$ profiles of the symbiotic star SY Mus obtained by van Winckel 
et al. (1993), and the template profiles. 
The profiles are enlarged by 5 times to reveal the details of the wing parts.
By the thick dashed lines we represent the profiles given by
the relation $f(\Delta v) \propto \Delta v^{-2}$, 
The dotted and long dashed lines show
the wing profiles that are generated using the Monte Carlo technique
adopted in the work of Lee \& Hyung(2000). The long dashed lines
represent the wing profiles obtained from the scattering region
of H~I column density $N_{HI} = 5\times 10^{19} {\rm\ cm^{-2}}$ and the
dotted lines from $N_{HI}= 10^{19}{\rm\ cm^{-2}}$.} 
\end{figure}

\begin{figure}
\plotone{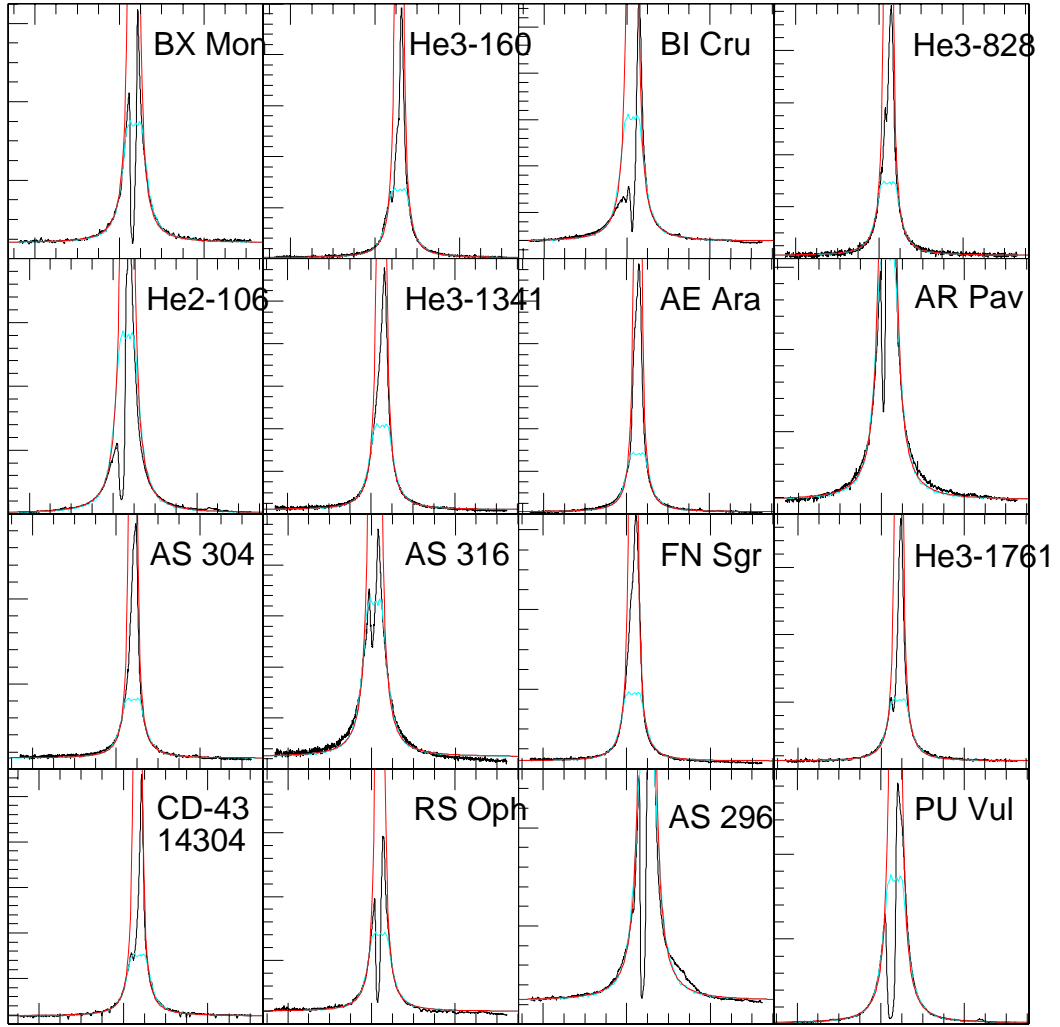}
\caption{
The H$\alpha$ profiles of 16
symbiotic stars selected from the atlas of H$\alpha$ spectra
compiled by van Winckel et al. (1993) and Ivison et al. (1994).
Objects with conspicuous wings were selected, and
no particular selection criterion was applied. 
The thick dashed lines and the dotted lines represent the same quantities 
as in Fig.~1, and
we omitted the case of $N_{HI} = 5\times 10^{19} {\rm\ cm^{-2}}$
for simplicity of presentation.
We omit the labels of the horizontal axis 
and vertical axis, which are the same as in Fig.~1.
The excellent agreement between the template wing profiles and the 
observed ones is apparent. }
\end{figure}

\end{document}